\documentclass{jaa}
\ExplSyntaxOn
\cs_set_eq:NN \year \tex_year:D
\ExplSyntaxOff
\usepackage{natbib}
\usepackage{graphicx} % for including images I think
\usepackage{xcolor} % for giving the text colour
\usepackage{hyperref} % for hyper ref of citations
\usepackage{gensymb} % for symbols
\usepackage{soul} % for strikethrough
\usepackage{siunitx} % for angstrom
\usepackage{csvsimple} % for .csv file
\usepackage{longtable} % for long tables
\usepackage{pgfplotstable} % for including tables or csv files
\usepackage{pdfpages} % for displaying PDFs
\usepackage{booktabs}
\usepackage{array}

\hypersetup{
    colorlinks=true,
    linkcolor=blue,
    citecolor=blue,
    urlcolor=blue
}
\begin{document}\sloppy

\title{TWIN-2: Timing Solar Radio Bursts between 180-350 MHz Against GOES and Aditya-L1 X-ray Detections}

\author{G N Paneendra\textsuperscript{1*}, Vinod K R\textsuperscript{1}, Arul Pandian B\textsuperscript{1}, Abhishek R\textsuperscript{1}, Susan Syeda\textsuperscript{1,2},
Chandana R\textsuperscript{2},
Dhanushree V\textsuperscript{2},
Ishwarya T\textsuperscript{2}, 
Manasa K N\textsuperscript{2}, 
Shiekh Saniya\textsuperscript{2}, 
Suresh A\textsuperscript{2},
Yashwanth R\textsuperscript{2},
Supriya Nayaka M K\textsuperscript{2} and
Prabu~T\textsuperscript{1}}

\affilOne{\textsuperscript{1}Electronics Engineering Group, Raman Research Institute, Bengaluru.\\\textsuperscript{2}Department of Physics, AES National Degree College, Gauribidanur.\\}

\twocolumn[{

\maketitle

\corres{paneendra.res@gmail.com}

%\msinfo{1 January 2015}{1 January 2015} % Submitted and accepted date

\begin{abstract}

We have conducted continuous solar observations for approximately two and half years using the Two-Element Interferometer (TWIN) at Gauribidanur Radio Observatory (RRI). During which several solar events have been detected. Our observational period coincides with the maximum events phase of the Solar Cycle 25, providing us opportunity to analyse the evolution of the solar activity. Our analysis of two and half years data show a clear increase in the occurrences of solar events, which is consistent with the expected rise in the solar activity towards the peak phase of the Solar Cycle 25. We further investigate the detected solar radio events using X-ray observations from the space based Aditya-L1 and the Geostationary Operational Environmental Satellites (GOES) satellites. For the automated detection of SRBs, a machine learning suite has been trained. This also serves as an instrument for space weather monitoring. On the 7th of November 2025, a solar storm was detected by our pipeline in realtime. A brief overview of the data-processing pipeline, along with the RFI mitigation techniques employed to identify and characterize the solar events, is also presented. Observation data of two and a half years of TWIN is available for public access. The data has been used in student training programmes on solar radio data analysis, with students carrying out the analysis and contributing to the initial analysis of the detected events.

%\hl{In the abstract, we can suitably mention about a) the  2 and a half year TWIN data availability in a portal. b) student training programs on data analysis c) efforts towards ML based detection, d) Implecations for space weather highlighting the Nov-7th detection and e) Future realtime detection plans across a wider frequency range}

%With Two-Element Interferometer (TWIN) at the Gauribidanur Radio Observatory we have been observing SUN for two and half years and have detected several solar events over this time. Our observation time is at the same time as the peak of Solar Cycle 25. A growth in solar events is seen in our data as it was expected during the peak of Solar Cycle 25 which arrives at the end of 2025. A comparison with the events seen in radio frequency and x-ray frequency is made where we observed that not all the events have same time of occurrence in both the frequencies. We have made an approximation of the how many events has their radio signatures as pre-cursors, post-cursors or simultaneous with x-ray signatures. Brief details on the data processing pipeline and RFI mitigation techniques are discussed.

\end{abstract}


}]

%%%%%%%%%%%%%%%%%%%%%%%

\section{Introduction}

Solar Radio Bursts (SRBs) are signatures of Solar flares and Coronal Mass Ejections in radio frequency. Solar flares are sudden releases of energy due to magnetic reconnection \citep{SF1}, \citep{SF2_SW1}. During solar flares, energy stored in magnetic fields is rapidly converted into thermal, kinetic and mechanical energies \citep{SF2_SW1} \textit{and is released into the solar system}. Radiation from these events are emitted across a wide range of the electromagnetic spectrum \citep{SF3}. Coronal mass ejections (CMEs) are huge expulsions of magnetized plasma from the sun \citep{cme1}. CME causes coronal shocks which is one of the mechanisms for SRB production. Majority of the radio waves emission from the sun is due to individual electrons being accelerated either because of collisions with ions (bremsstrahlung) or because of spiraling in a magnetic field (gyro-synchrotron emission) \citep{REM}. CMEs have a major influence on Space Weather and the near-Earth conditions \citep{SF2_SW1}. 

SRB are classified into five types based on frequency bandwidth, frequency drift rate, and duration of occurrence which are commonly associated with active sun. Table \ref{tof} presents an overview of the it \citep{SBC1}.

\section{Radio Sun}

The Sun is the considered to be a blackbody and the temperature of the sun can be estimated by plotting a blackbody curve based on the optical observation and the resultant temperature is 6000\degree k. But if we plot a blackbody curve based on the radio frequency observation, the blackbody curve fit can be seen going from 100,000\degree K to 10\textsuperscript{10}\degree k \citep{radioastronomybykraus}. If the sun was radiating only as a thermal source then the variation in the radio flux shouldn't exist. 

Non-thermal emission from the sun is caused by variety of source; Langmuir wave which are cause by anisotropic electrons traveling in the magnetic field lines, mildly relativistic particles emitting gyrosynchroton radiation \citep{Nindos},\citep{nont} and relativistic emitting incoherent synchrotron radiation \citep{nont}. The non-thermal emission from the corona and upper corona emit radio waves within the frequency of 180 - 350 MHz. Type III, IV, V and Type II \citep{SF4} SRBs are know to be visible the mentioned band width. The radio sun's diameter is know to change at different radio frequency and during different stages of solar cycle \citep{radiosundia1}.

\begin{table}[h]
    \centering
    \renewcommand{\arraystretch}{1.5}
    \resizebox{0.49\textwidth}{!}{
    \begin{tabular}{|c|c|c|c|c|}
        \hline
        \textbf{TYPE} & \textbf{FREQUENCY RANGE} & \textbf{MINIMUM} & \textbf{MAXIMUM} & \textbf{TYPICAL} \\ \hline
        Type I    & 80 - 200 MHz     & 0.1 s     & 1 Week     & Hours-Days     \\ \hline
        Type II    & 20 - 150 MHz     & 10 min     & 90 min     & 3 - 30 min     \\ \hline
        Type III    & 10 KHz - 1 GHz     & 0.1 s     & 33 min     & 1 - 5 min     \\ \hline
        Type IV    & 20 MHz - 2 GHz     & 10 min     & 9 hr     & 45 min - 2 hr     \\ \hline
        Type V   & 10 - 200 MHz     & Few sec      & Data     & 1 - 3 min     \\ \hline 
    \end{tabular}
    }
    \label{tof}
    \caption{Classification of Solar Radio Bursts based on the frequency bandwidth and duration of occurrence \citep{Paneendra2025}}
\end{table}

%The non-thermal emission from the corona and upper corona are visible in our bandwidth which includes Type III, IV and V \citep{SF4}. We have also observed high frequency parts of the Type II bursts occasionally. The radio sun's diameter changes over the radio frequency and during different stages of solar cycle \citep{radiosundia1}.

Solar radio emissions at the 180 - 350 MHz are dominated by coherent plasma emission at the local plasma frequency or its harmonics, making this range highly sensitive to fast electron beams, CME driven shocks, and turbulent coronal structures. Even a small number of accelerated electrons generate strong, high-brightness radio bursts in this bandwidth \citep{ReidRatcliffe2014}. The electron densities corresponding to 180 - 350 MHz map to heliocentric heights of roughly 1.2 - 1.6 solar radius, the same coronal layers where CME driven shocks form and flare-accelerated electrons begin to escape into the heliosphere, making this band a direct probe of the earliest phases of eruptive solar activity \citep{Gopalswamy2013}, \citep{White2024}. SRBs coming under Type II have a duration of less than a minute, but when observed over a wider frequency bandwidth, they can persist for longer duration \textit{(but not too long)}. Similarly, other types of SRBs, Type II and Type V, have a duration of several minutes, while Type I and Type IV can occur for several hours \citep{Paneendra2025}. Observing SRBs in a broad frequency bandwidth will help detect thermal and non-thermal radiation. 

Our observation at 180 - 350 MHz (1.6 - 0.8 m) is made using Two-Element Interferometer (TWIN) located at Gauribidanur Radio Observatory, RRI (Longitude: 77.428° E, Latitude: 13.603° N) \citep{TWIN1}. Our observing frequency comes in places where the ionospheric distortions are not dominated (below 100 MHz it is dominated) and plasma emission are not suppressed (it is above 500 MHz approx). Giving us a clear window for tracking flare \textit{(active sun)} related particle acceleration and CME onsets \citep{ReidRatcliffe2014}, \citep{deGasperin2018}. The observations are made everyday typically for 6 hours. It is a transient mode observation. The data is recorder in burst mode with 3 second interval sampled at 800 ms or 1.25 GS/s with ADC resolution of 8 bit\citep{TWIN1}. 

\begin{figure}[h]
    \centering
    \includegraphics[width=0.9\linewidth]{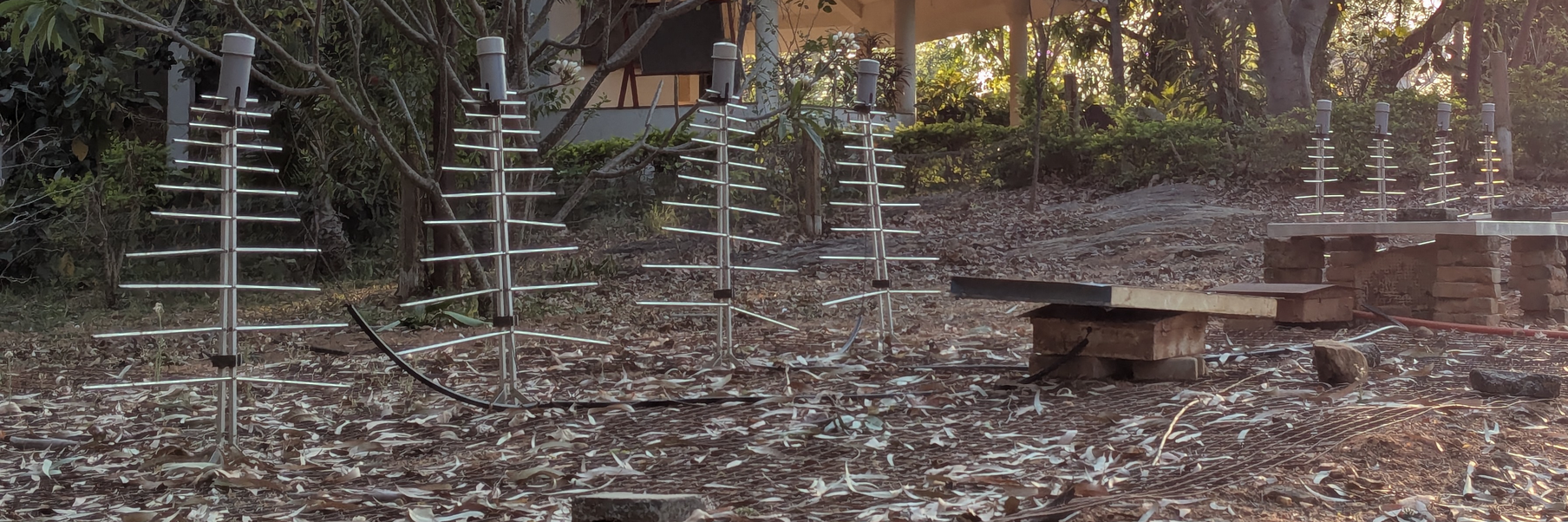}
    \caption{Two-Element Radio Interferometer (TWIN) at Gauribidanur Radio Observatory (RRI), consisting of 8 LPDAs, combining 4 LPDAs into one element each with a baseline of about 9m between two elements.}
    \label{twinimage}
\end{figure}

\section{Analysis (Pre-processing of data)}

The information in the data (signal from a radio antenna) can be represented as \cite{RFI1}:
\begin{equation}
    x(t) = x_{sig}(t)+x_{sys}(t)+x_{RFI}(t)
    \label{madeqn1}
\end{equation}
Where:

$x_{sig}(t)=$ Signal of interest(astronomical signal)

$x_{sys}(t)=$ System noise (undesired noise)

$x_{RFI}(t)=$ Radio Frequency Interference (undesired signal)\\

$x_{sig}(t)$ is the required part of the collected signal. Initial data processing involves taking the raw ADC values and performing mean subtracting. In our observation data, the RFI is from both terrestrial and satellite sources. The RFI mitigation in the time series carries out by MAD \citep{mad} based RFI mitigation. A sliding window of 512 points is created which has a step size of 1 and it goes through all the data points, median and MAD is calculated for each window. We have defines two condition:

\begin{equation}
median \pm K \times MAD
\label{eqn2}
\end{equation}

\begin{figure*}[!ht]
    \centering

    \begin{minipage}{0.48\textwidth}
        \centering
        \includegraphics[width=\textwidth]{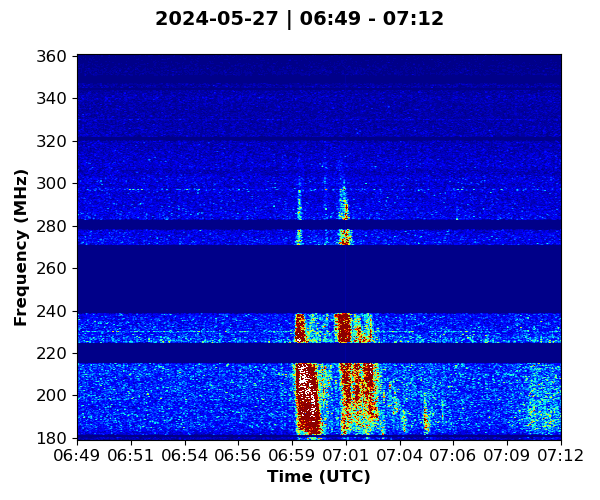}       
        \textbf{(a) Type II}
        \label{fig:type2}
    \end{minipage}
    \hfill
    \begin{minipage}{0.48\textwidth}
        \centering
        \includegraphics[width=\textwidth]{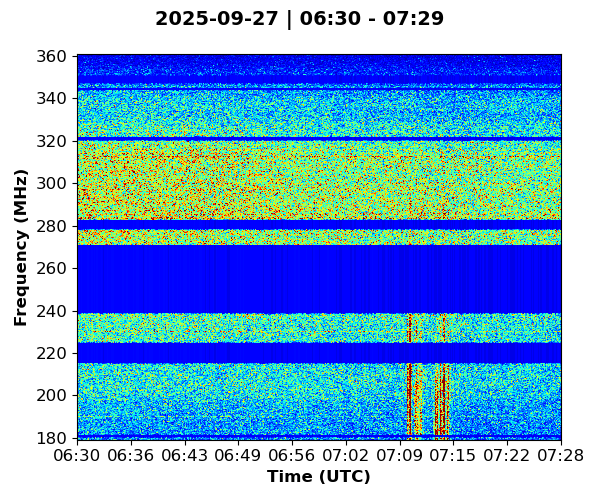}
        \textbf{(b) Type III}
        \label{fig:type2}
    \end{minipage}

    \vspace{0.25cm}

    \begin{minipage}{0.48\textwidth}
        \centering
        \includegraphics[width=\textwidth]{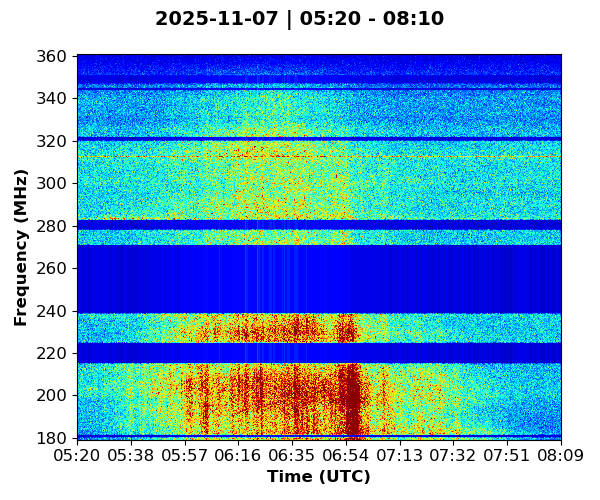}
        \textbf{(c) Type IV}
        \label{fig:type4}
    \end{minipage}
    \hfill
    \begin{minipage}{0.48\textwidth}
        \centering
        \includegraphics[width=\textwidth]{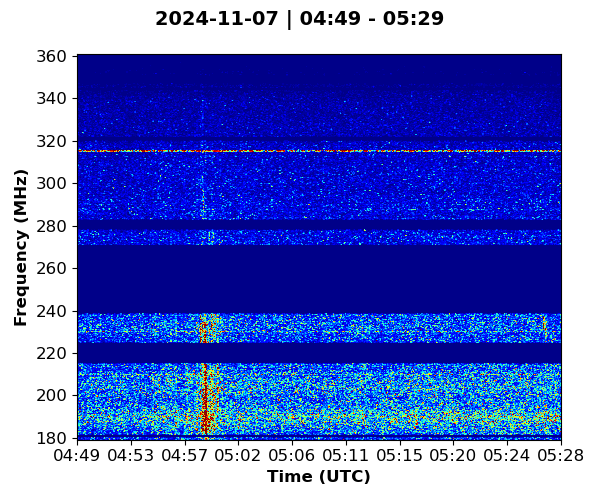}
        \textbf{(d) Type V}
        \label{fig:type5}
    \end{minipage}

    \caption{Figure (a) shows a Type II SRB occurred on 27 May 2024 from 06:59 till 07:06 UT. Figure (b) shows a group of Type III SRB occurred on 27 September 2025 from 07:09 till 07:15 UT. Figure (c) shows a Type IV SRB occurred on 07 November 2025 from 05:30 till 05:50 UT. Figure (d) shows a Type V SRB occurred on 07 November 2024 between 04:58 and 05:00 UT.}
    \label{fig:four_types}
\end{figure*}

Where K is threshold multiplier given for our requirement. The data points which fall into the conditions \ref{eqn2} are masked and replaced with the median of that window. This process is compute expensive and requires large time for processing. To minimize the time, MAD based RFI mitigation process has been parallel run on the CPU. The algorithm before parallelization used to take \texttildelow55 minutes and after takes \texttildelow5 minutes. The mentioned RFI mitigation technique may to be required for all the days of observation. After the time domain RFI mitigation, FFT is performed to obtain the frequency information and the known satellite RFIs is mitigated in the frequency domain. And cross-correlation of signal from two elements.

\section{Detection of SRBs}

RFI-cleaned and averaged spectrographs were used to identify SRB signatures. We have employed a semi-automated analysis to locate the signatures and epochs. Currently, we have about 250 possible detections from the two-and-a-half-year observation. The detections include signatures from solar storms, CMEs, and Type II. Type III, Type IV and Type V SRBs. A total of 103 events are analysed and present in Table \ref{tab:event_types}. Figure \ref{fig:four_types} presents our detection of Type II, III, IV and V SRBs.

\begin{table}[h]
    \centering
    \begin{tabular}{|c|c|}
        \hline
        \textbf{Event Type} & \textbf{Number of Events} \\
        \hline
        Type II & 7 \\ \hline
        Type III & 89 \\ \hline
        Type IV & 4 \\ \hline
        Type V & 8 \\ \hline
        Complex & 15 \\ \hline
        UC (Un-categorized) & 7 \\
        \hline
        Total & 123 \\
        \hline
    \end{tabular}
    \caption{The table shows the total number of SRBs signatures successfully categorized and the number of SRBs belonging to each category.}
    \label{tab:event_types}
\end{table}

\begin{figure}[h]
    \centering
    \includegraphics[width=0.9\linewidth]{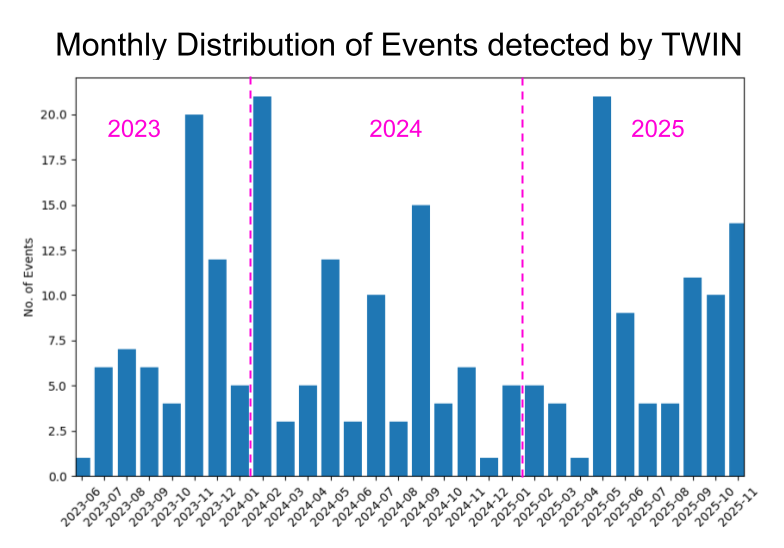}
    \caption{The number of events recorded per month and the events increase by the end of 2025 which is the peak solar activity time of solar cycle 25.}
    \label{solaractivity1}
\end{figure}

\begin{figure*}[h!]
\centering

% Row 1

\begin{minipage}{0.19\textwidth}
\centering
\includegraphics[width=\linewidth]{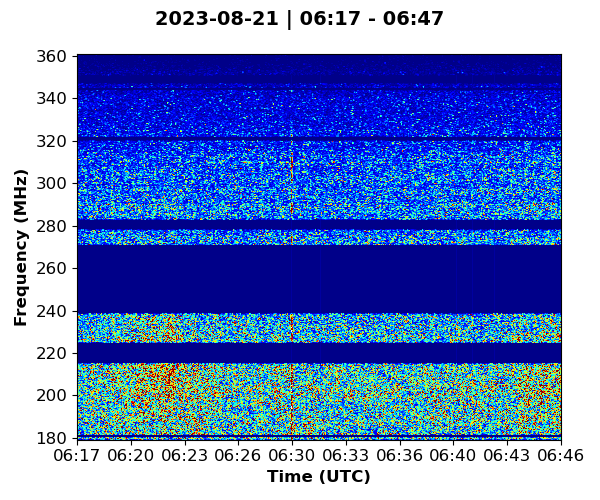}
\par\vspace{2pt}(a)
\end{minipage}\hfill
\begin{minipage}{0.19\textwidth}
\centering
\includegraphics[width=\linewidth]{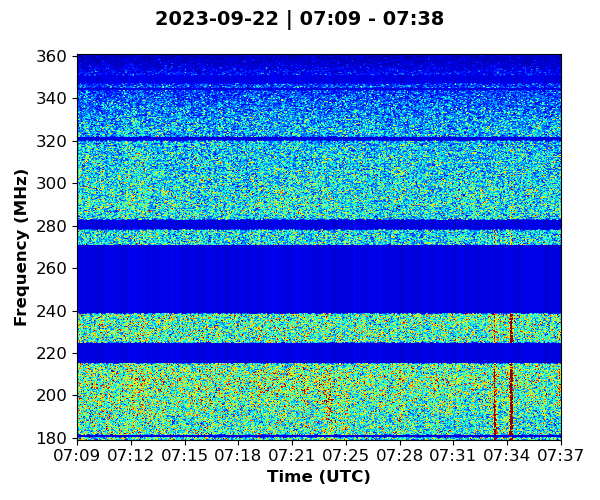}
\par\vspace{2pt}(b)
\end{minipage}\hfill
\begin{minipage}{0.19\textwidth}
\centering
\includegraphics[width=\linewidth]{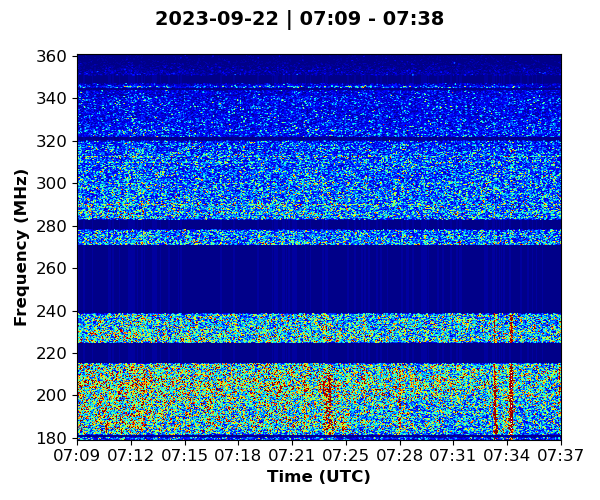}
\par\vspace{2pt}(c)
\end{minipage}\hfill
\begin{minipage}{0.19\textwidth}
\centering
\includegraphics[width=\linewidth]{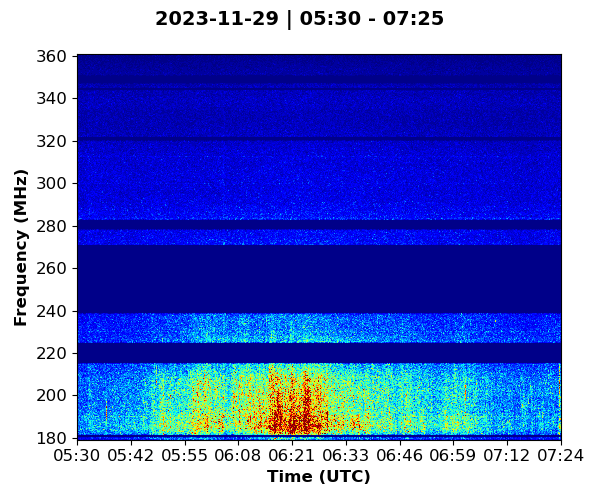}
\par\vspace{2pt}(d)
\end{minipage}\hfill
\begin{minipage}{0.19\textwidth}
\centering
\includegraphics[width=\linewidth]{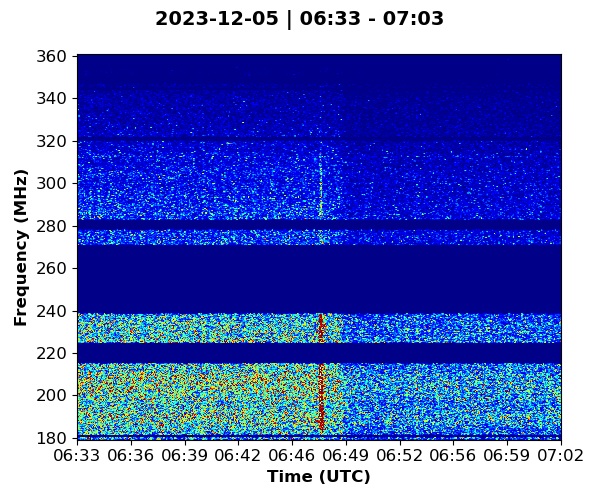}
\par\vspace{2pt}(e)
\end{minipage}

\vspace{6pt}

% Row 2

\begin{minipage}{0.19\textwidth}
\centering
\includegraphics[width=\linewidth]{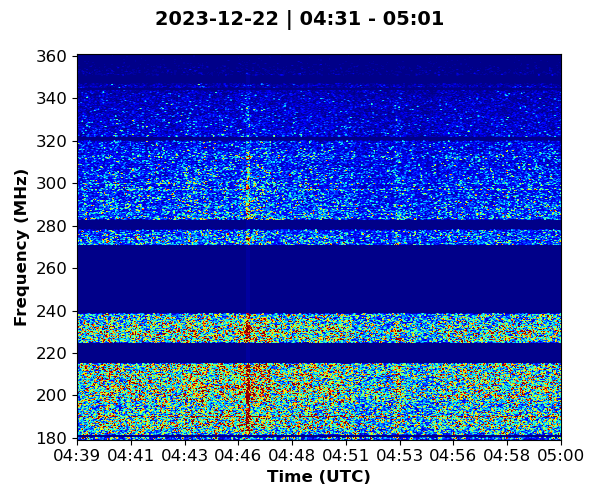}
\par\vspace{2pt}(f)
\end{minipage}\hfill
\begin{minipage}{0.19\textwidth}
\centering
\includegraphics[width=\linewidth]{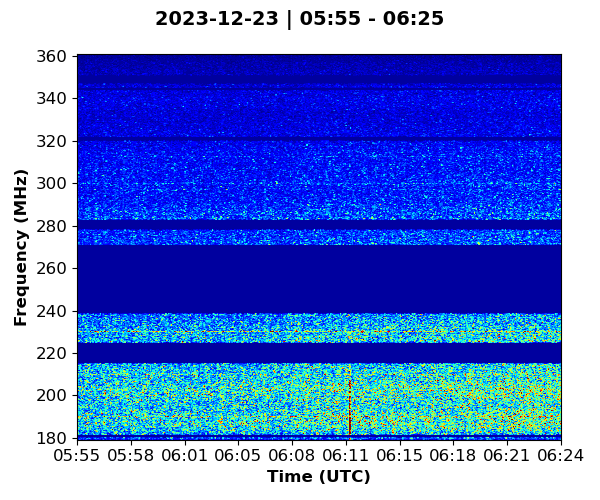}
\par\vspace{2pt}(g)
\end{minipage}\hfill
\begin{minipage}{0.19\textwidth}
\centering
\includegraphics[width=\linewidth]{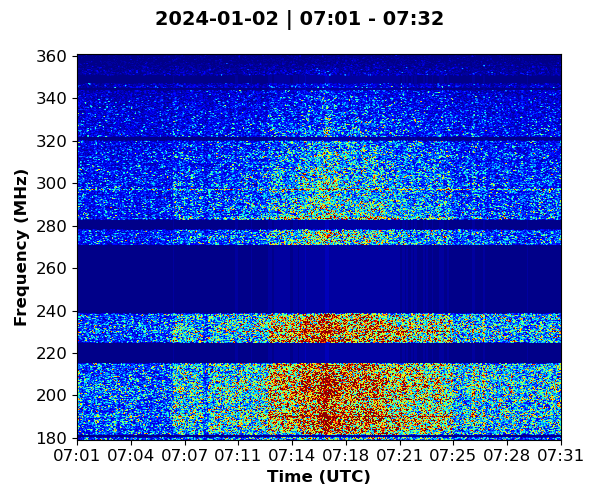}
\par\vspace{2pt}(h)
\end{minipage}\hfill
\begin{minipage}{0.19\textwidth}
\centering
\includegraphics[width=\linewidth]{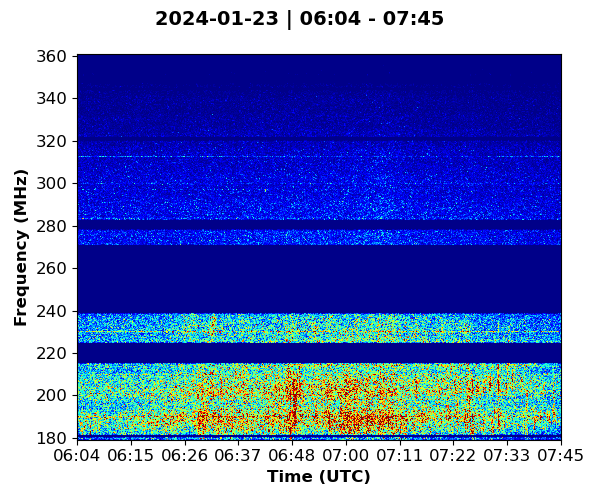}
\par\vspace{2pt}(i)
\end{minipage}\hfill
\begin{minipage}{0.19\textwidth}
\centering
\includegraphics[width=\linewidth]{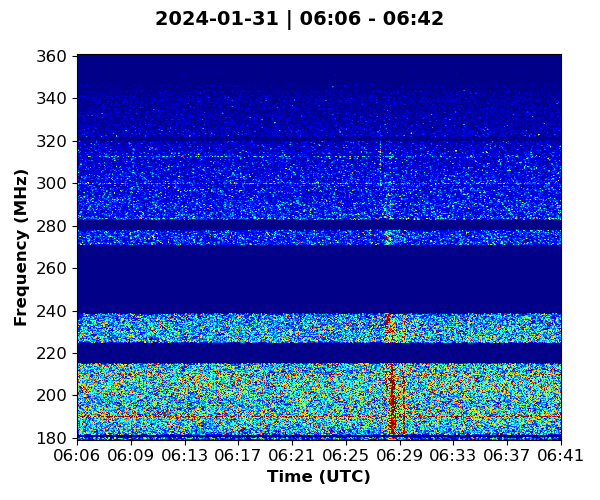}
\par\vspace{2pt}(j)
\end{minipage}

\vspace{6pt}

% Row 3

\begin{minipage}{0.19\textwidth}
\centering
\includegraphics[width=\linewidth]{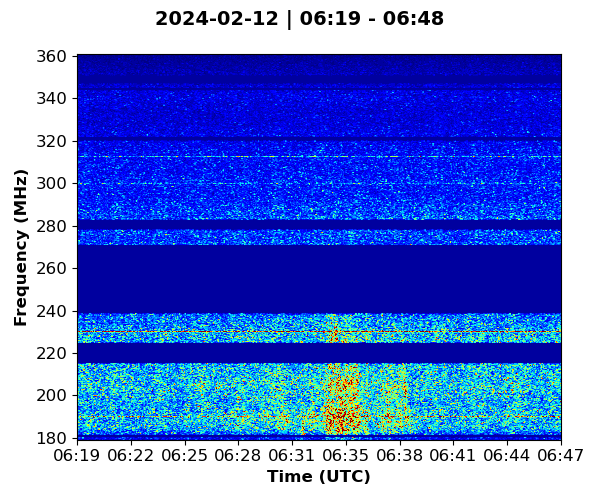}
\par\vspace{2pt}(k)
\end{minipage}\hfill
\begin{minipage}{0.19\textwidth}
\centering
\includegraphics[width=\linewidth]{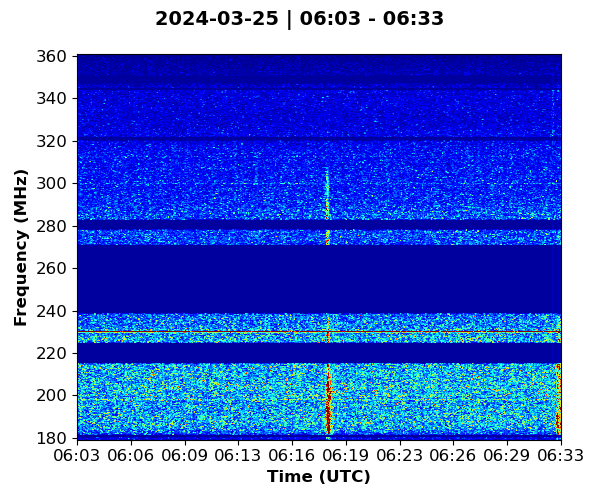}
\par\vspace{2pt}(l)
\end{minipage}\hfill
\begin{minipage}{0.19\textwidth}
\centering
\includegraphics[width=\linewidth]{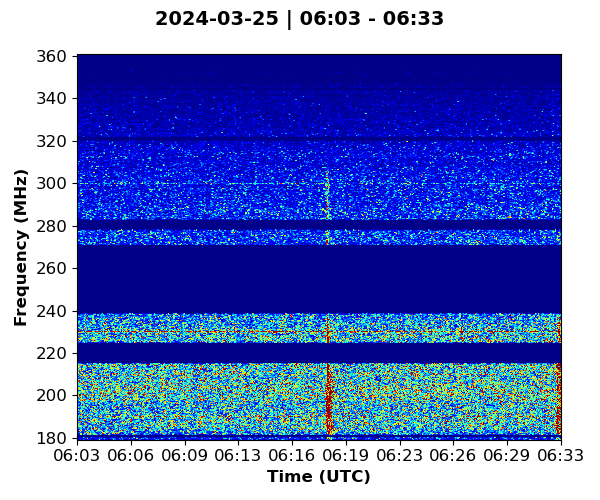}
\par\vspace{2pt}(m)
\end{minipage}\hfill
\begin{minipage}{0.19\textwidth}
\centering
\includegraphics[width=\linewidth]{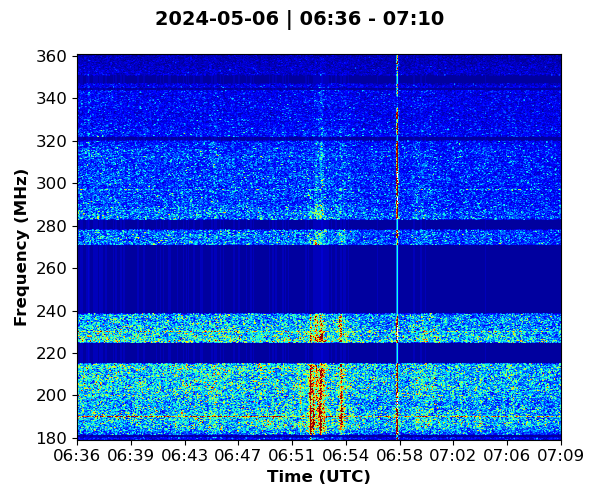}
\par\vspace{2pt}(n)
\end{minipage}\hfill
\begin{minipage}{0.19\textwidth}
\centering
\includegraphics[width=\linewidth]{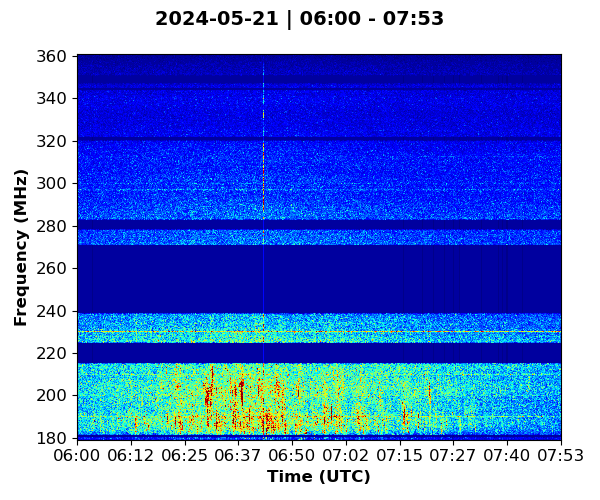}
\par\vspace{2pt}(o)
\end{minipage}

\vspace{6pt}

% Row 4

\begin{minipage}{0.19\textwidth}
\centering
\includegraphics[width=\linewidth]{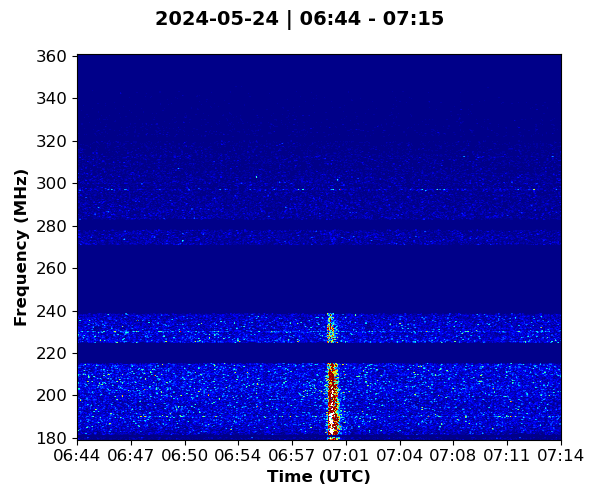}
\par\vspace{2pt}(p)
\end{minipage}\hfill
\begin{minipage}{0.19\textwidth}
\centering
\includegraphics[width=\linewidth]{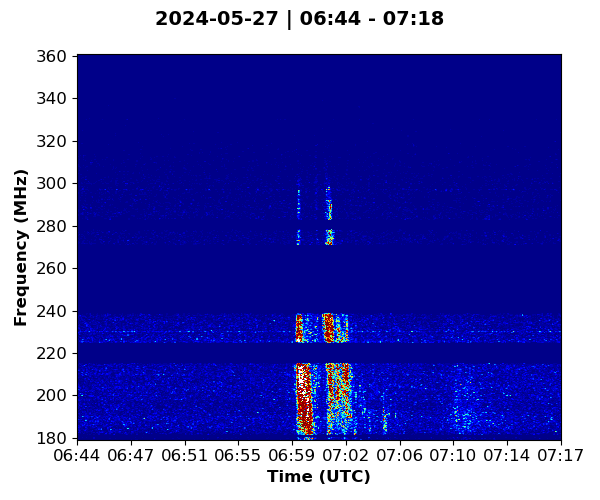}
\par\vspace{2pt}(q)
\end{minipage}\hfill
\begin{minipage}{0.19\textwidth}
\centering
\includegraphics[width=\linewidth]{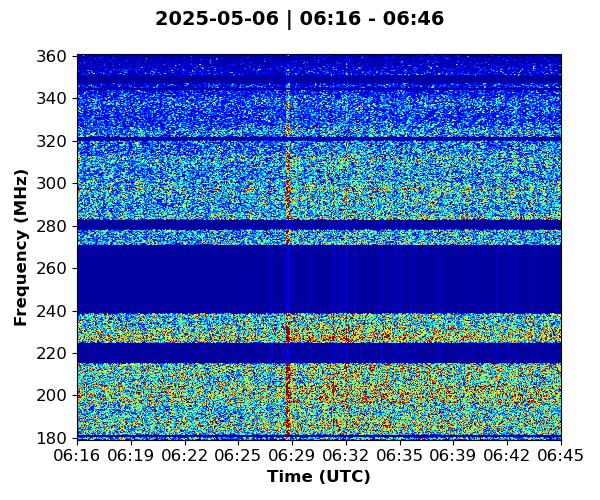}
\par\vspace{2pt}(r)
\end{minipage}\hfill
\begin{minipage}{0.19\textwidth}
\centering
\includegraphics[width=\linewidth]{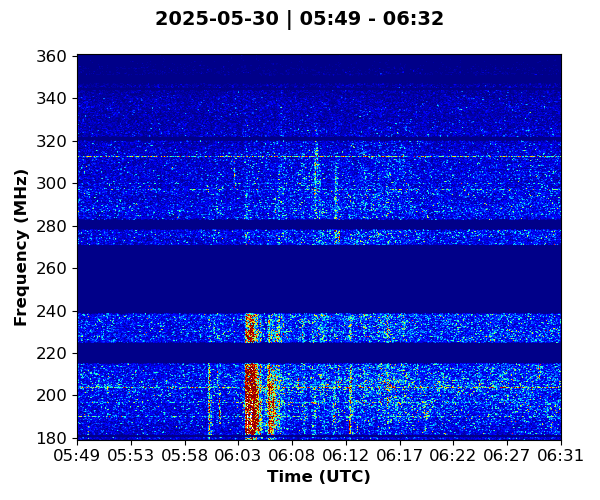}
\par\vspace{2pt}(s)
\end{minipage}\hfill
\begin{minipage}{0.19\textwidth}
\centering
\includegraphics[width=\linewidth]{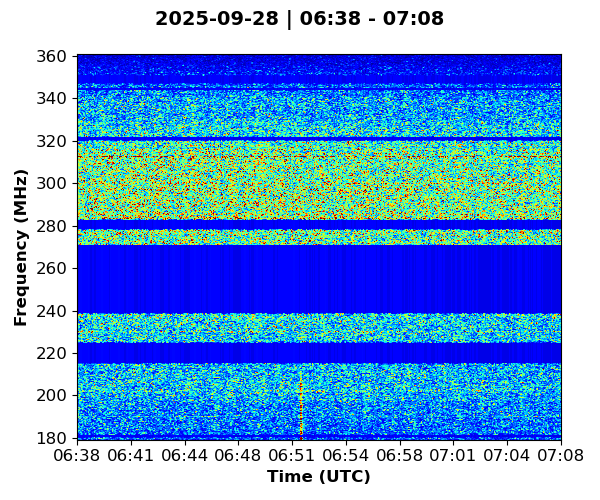}
\par\vspace{2pt}(t)
\end{minipage}

\vspace{6pt}

% Row 5

\begin{minipage}{0.19\textwidth}
\centering
\includegraphics[width=\linewidth]{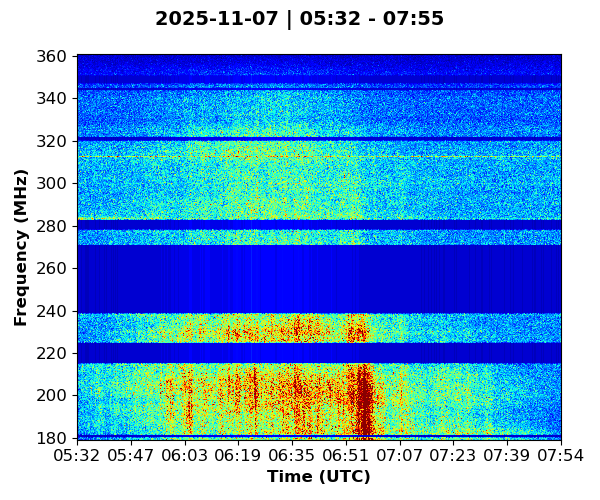}
\par\vspace{2pt}(u)
\end{minipage}\hfill
\begin{minipage}{0.19\textwidth}
\centering
\includegraphics[width=\linewidth]{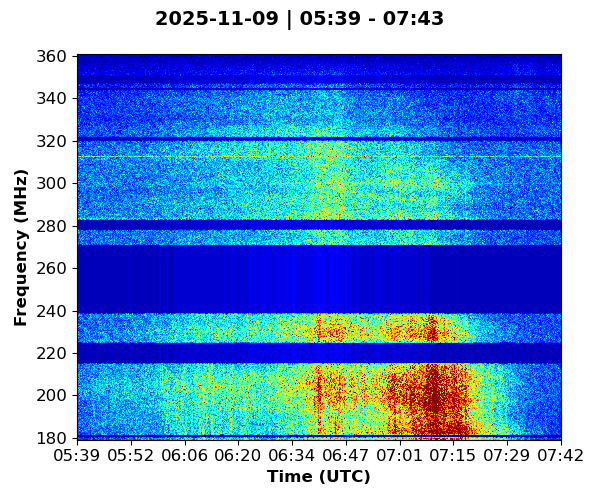}
\par\vspace{2pt}(v)
\end{minipage}\hfill
\begin{minipage}{0.19\textwidth}
\centering
\includegraphics[width=\linewidth]{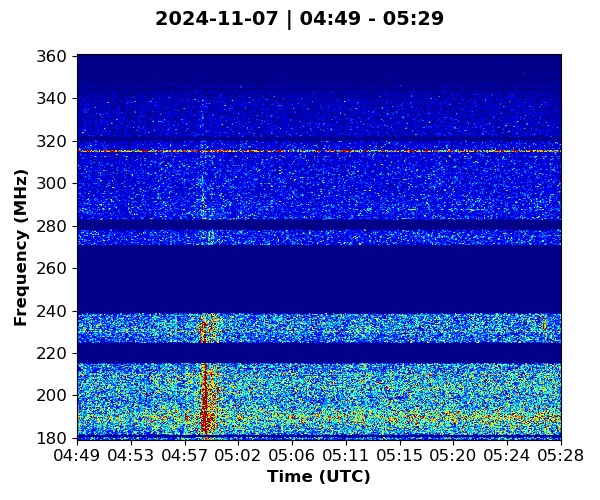}
\par\vspace{2pt}(w)
\end{minipage}\hfill
\begin{minipage}{0.19\textwidth}
\centering
\includegraphics[width=\linewidth]{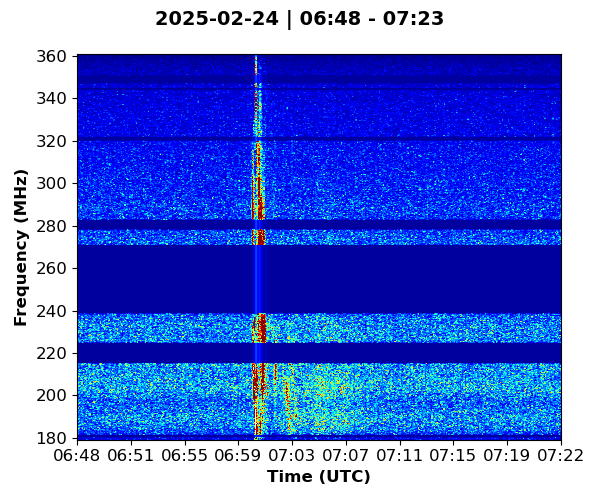}
\par\vspace{2pt}(x)
\end{minipage}\hfill
\begin{minipage}{0.19\textwidth}
\centering
\includegraphics[width=\linewidth]{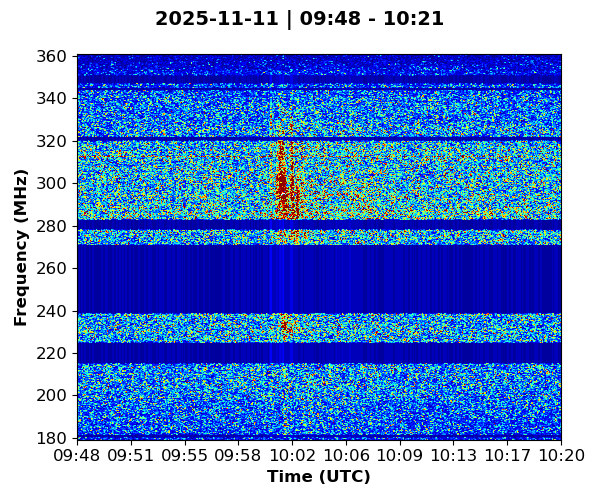}
\par\vspace{2pt}(y)
\end{minipage}

\caption{Spectrograph of solar radio bursts (SRBs) observed with TWIN. Images (a), (b), (c), (e), (f), (g), (j), (l), (m), (n) and (r) correspond to Type III SRBs; images (d), (h), (i), (o), (u) and (v) correspond to Type IV SRBs; images (k), (p), (q), (s), (t), (x) and (y) correspond to Type II SRBs; and image (w) corresponds to a Type V SRB.}
\label{fig:solar_radio_bursts}

\end{figure*}

We observed an increase in the number of events towards the end of 2025, possibly influenced by the intense activities of solar cycle 25. Figure \ref{solaractivity1} depicts the observed trend of solar activity over our observation periods between June 2023 and November 2025.  The number of Type-III SRBs is large in number. Type-II possibly occurred below our observation band, resulting in the detection of only their harmonics in our data, which explains the detection trend shown in Figure \ref{fig:dist_types}. In total, we have 103 days of confirmed Solar events out of 482 observation days. A selected gallery of solar bursts detected in our data is presented in Figure \ref{fig:solar_radio_bursts}.

\section{Comparing SRBs with x-ray flares}

A single eruptive event can radiate across the electromagnetic spectrum, so an SRB seen by TWIN may have an X-ray counterpart. We compared the start, peak and end times of each SRB with those of its X-ray flare, using GOES and HEL1OS (Aditya L1). GOES records two channels, 1.5--12.5~keV (1--8~\AA) and 3.1--24.8~keV (0.5--4~\AA). We used HEL1OS light curves in the 5--20, 20--30, 30--40, 40--60 and 1.8--90~keV bands.

\begin{figure}[!hb]
    \centering
    \includegraphics[width=0.9\linewidth]{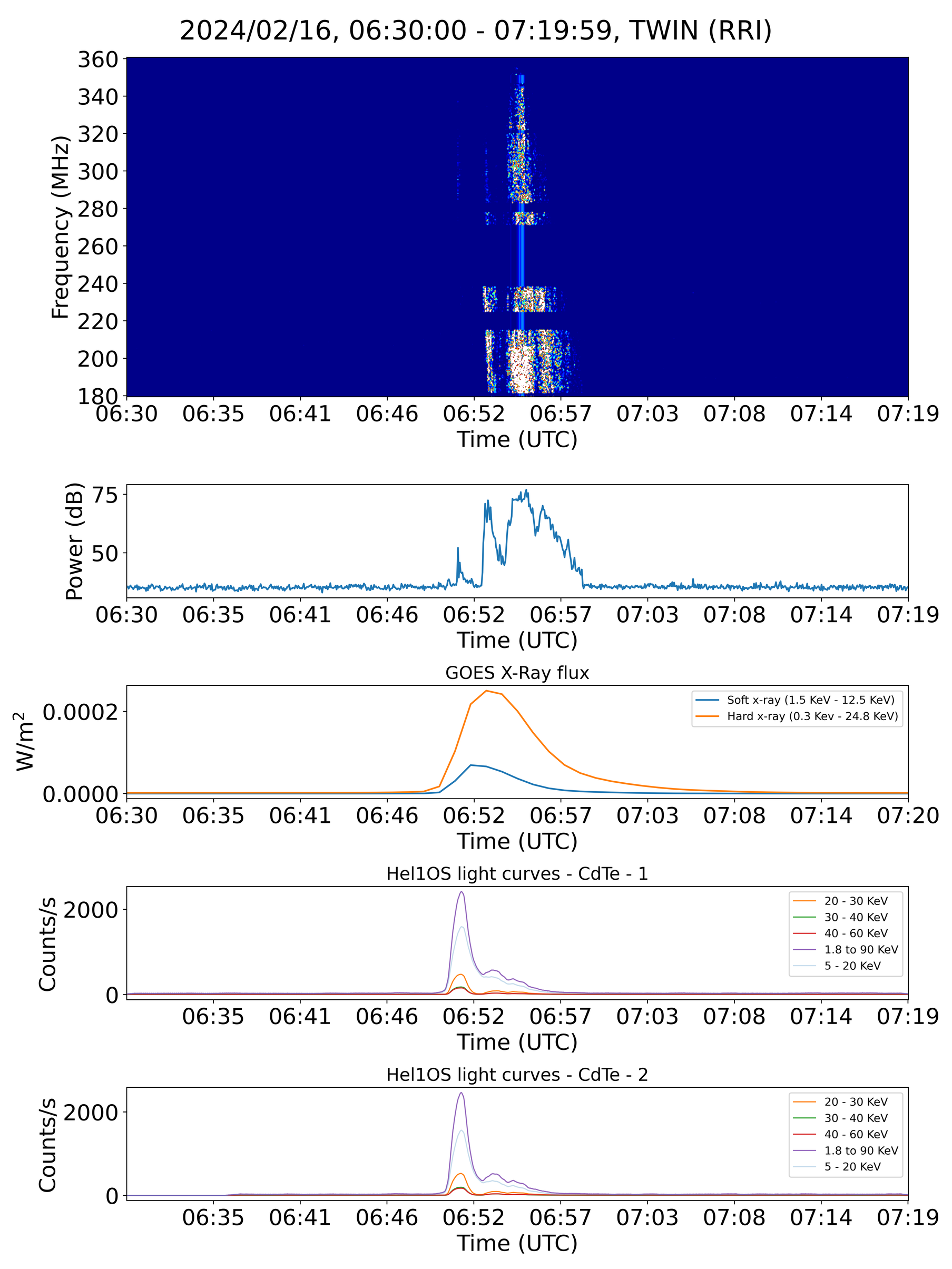}
    \caption{Type II SRB seen on 16 Feb 2024 with an X class (X2.47) flares. The events appears to have started from 06:50 UT and end at 06:58 UT. Where in the hard x-ray from GOES, the same events is visible from 06:50 UT and 06:57 UT, soft x-ray begins as the same time as hard x-ray but ends much later at 07:04 UT. The x-ray from Aditya L1 has the same event time as hard x-ray from GOES.}
    \label{type2_comp}
\end{figure}

\begin{figure}[h!]
    \centering
    \includegraphics[width=0.9\linewidth]{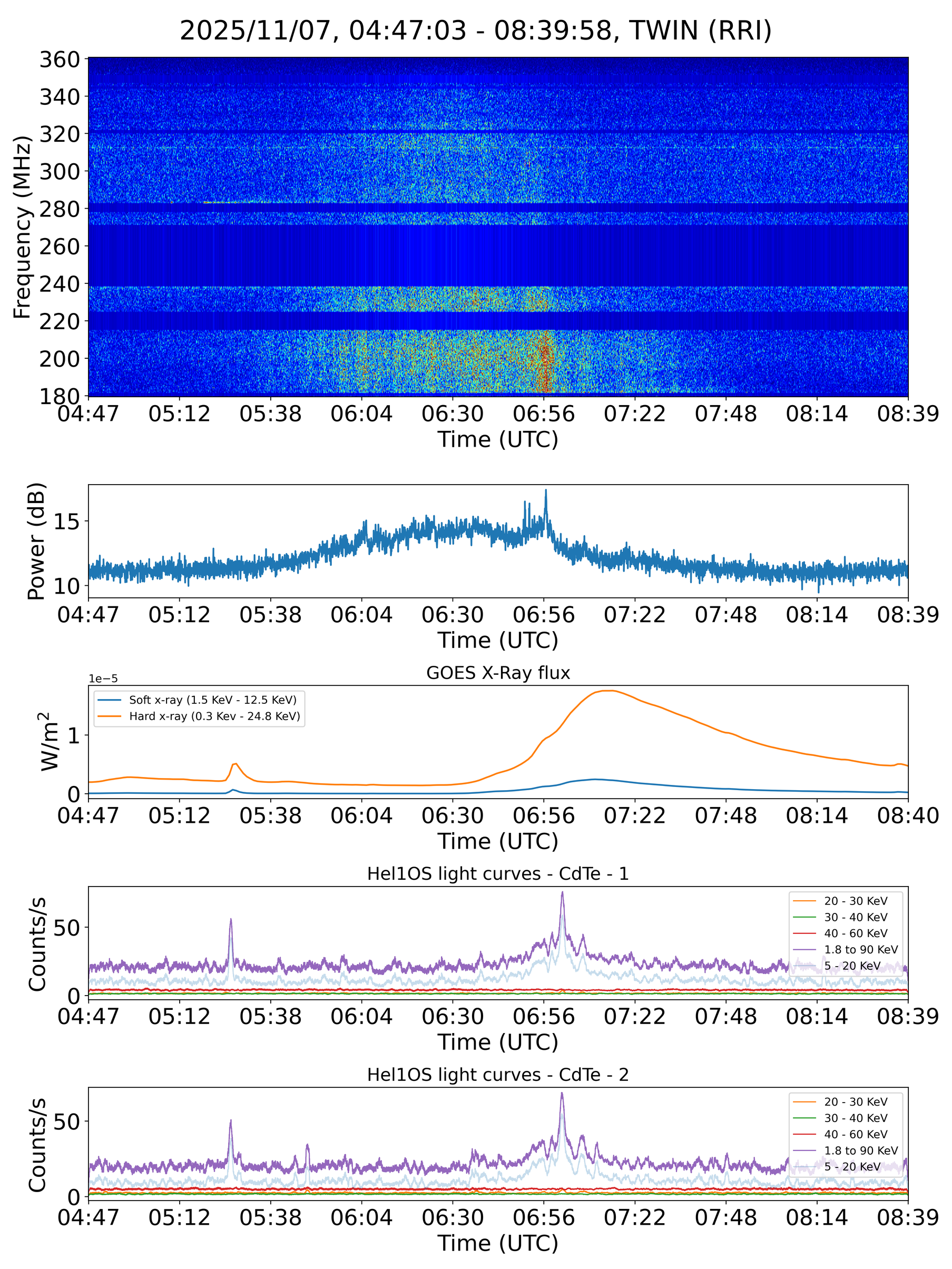}
    \caption{Type IV SRB observed simultaneously alongside an M class x-ray flares on 07 November 2025 by GOES satellite and HEL1OS. This event is appearing in the radio frequency from 05:30 UT till 07:50 UT, in x-ray the same events is visible from 06:30 UT and end beyond 10:00 UT according to GOES, where as in HEL1OS data, the event ends at 07:15 UT.}
    \label{type4_comp}
\end{figure}

\begin{figure}[h!]
    \centering
    \includegraphics[width=0.9\linewidth]{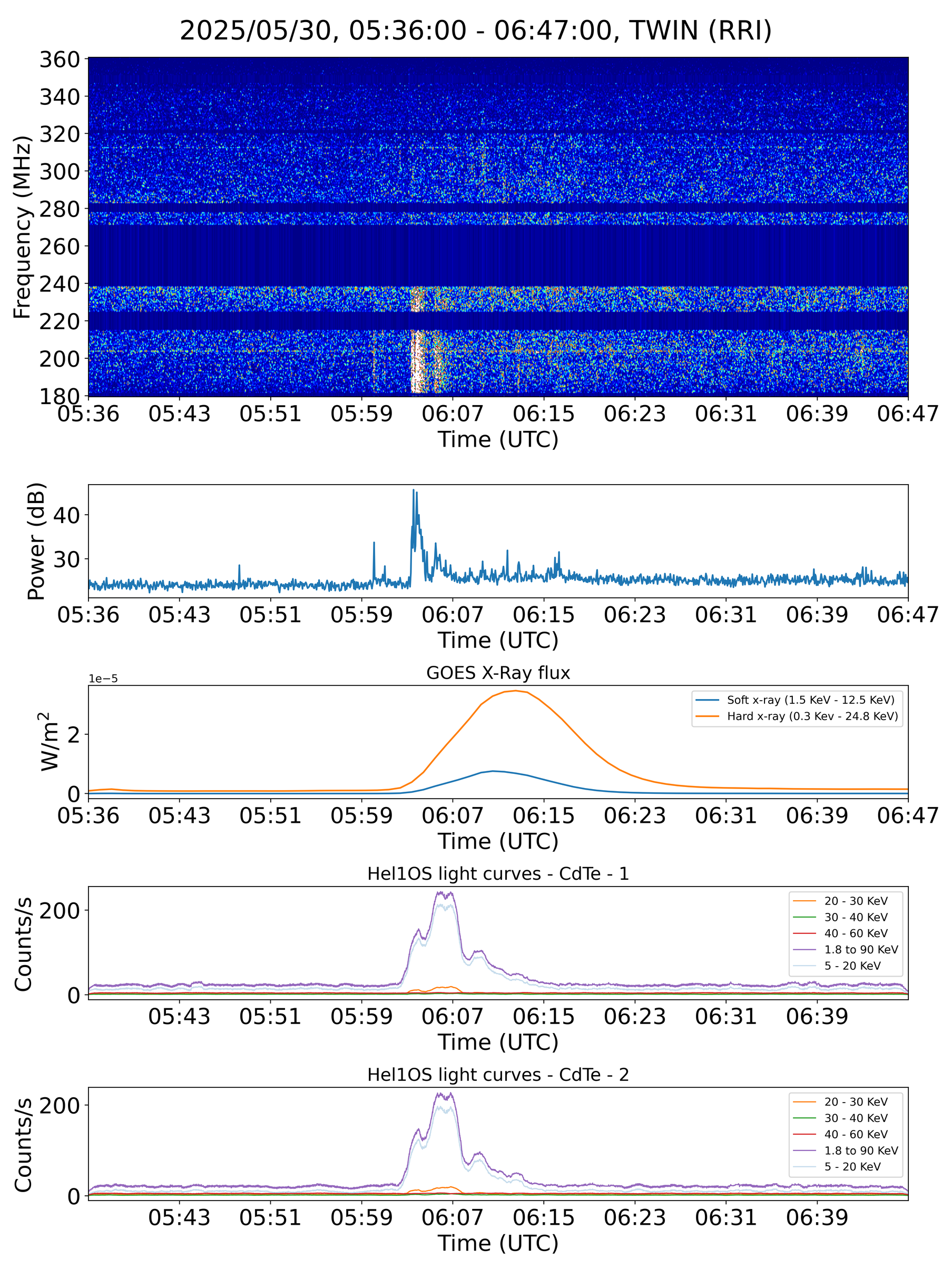}
    \caption{Type V SRB seen simultaneously alongside An M class flares on 30 May 2025. The Type V SRB is followed by a few Type IIIs SRBs. The hard x-ray signature for this event from GOES appears to have started from 06:03 UT and end at 06:21. Soft x-ray signature from GOES has started from 06:00 UT and ends at 06:28. The x-ray (hard) from the HEL1OS has a complex shape and start at 06:03 UT and ends at 06:18 UT.}
    \label{type5_comp}
\end{figure}

Figures~\ref{type2_comp}, \ref{type4_comp} and \ref{type5_comp} show one example of each case: the X-ray flare preceding the radio burst (Figure~\ref{type2_comp}), following it (Figure~\ref{type4_comp}), and occurring nearly simultaneously with it (Figure~\ref{type5_comp}). Each figure has five panels: the TWIN spectrograph, the band-averaged TWIN flux, the GOES X-ray flux, and the two HEL1OS (CdTe-1 and CdTe-2) light curve.

\begin{figure*}[h]
    \centering
    \includegraphics[width=\linewidth]{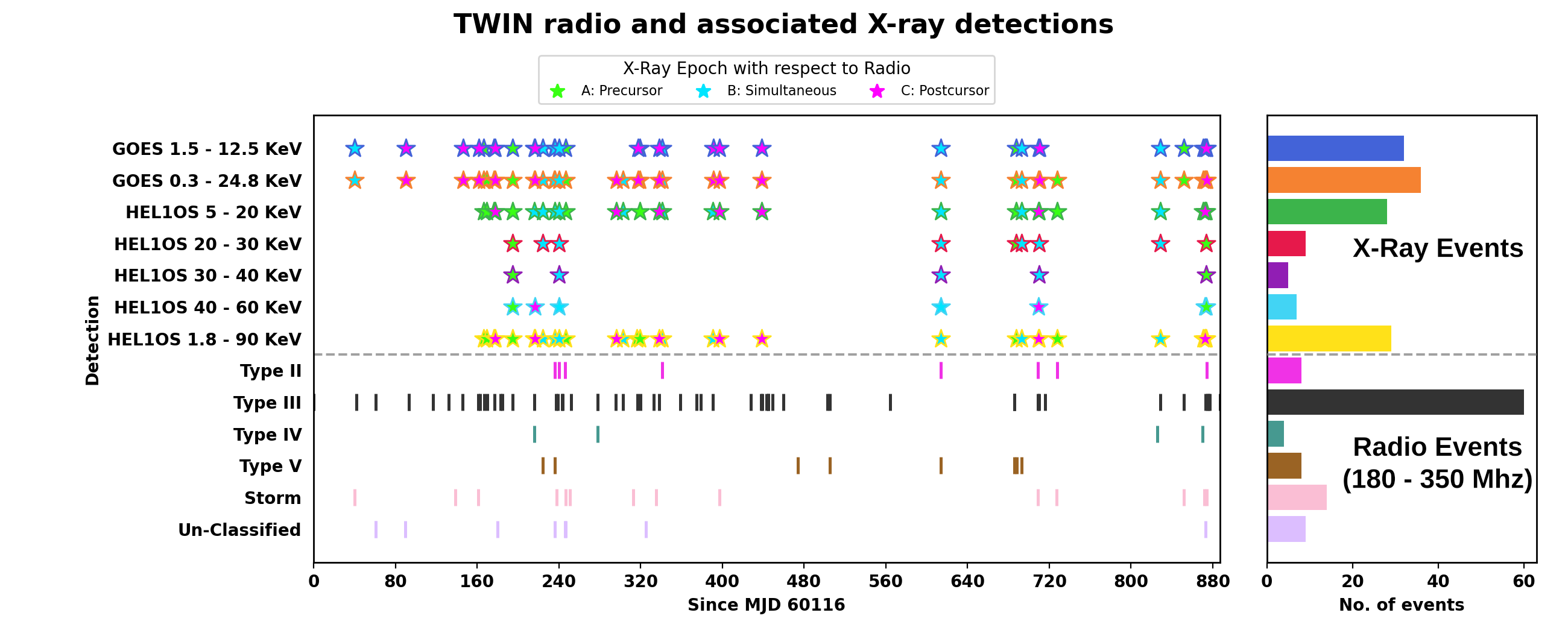}
    \caption{Distribution of Solar Radio Bursts occurred in the last 2 and half years over their Types and possible x-ray events associated with them.}
    \label{fig:dist_types}
\end{figure*}

We found X-ray counterparts for 45 events. In 13 events the X-ray flare leads the radio burst, 14 are coincident, and in 18 the X-ray flare lags the radio burst. Figure~\ref{fig:dist_types} shows their distribution over the observing period.

\textbf{X-ray precedes radio.} Figure~\ref{type2_comp} shows a Type II burst, with fundamental and harmonic bands and a slow frequency drift, on 16 February 2024 during an X2.47 flare. The radio burst and the X-ray flare begin together at about 06:50~UT, and all the emissions overlap in time. The HEL1OS hard X-rays peak first, and the radio flux peaks about 1~min later. The GOES flux peaks after HEL1OS. The radio burst ends at 06:58~UT. The HEL1OS emission ends earlier than that of GOES, close to the end of the radio burst, whereas the GOES flux decays more slowly, ending at 06:57~UT in the short-wavelength channel and 07:04~UT in the long-wavelength channel. Hard X-rays peaking before, and ending earlier than, the soft X-rays is the behaviour expected from the Neupert effect, in which the soft X-ray flux follows the time integral of the hard X-ray flux \citep{Neupert1968, DennisZarro1993}.

\textbf{X-ray follows radio.} Figure~\ref{type4_comp} shows a Type IV continuum on 7 November 2025, observed alongside an M-class flare. The radio emission lasts from 05:30 to 07:50~UT, whereas the X-ray flare starts at about 06:30~UT and continues beyond 10:00~UT in GOES (until 09:42~UT in HEL1OS). The X-ray flux peaks about 1~h after the first radio peak and 2.5~h after the second, and the flare outlasts the radio burst by more than two hours. Long-duration Type IV continua such as this are commonly associated with CMEs.

\textbf{X-ray and radio nearly simultaneous.} Figure~\ref{type5_comp} shows a Type V burst on 30 May 2025, observed alongside an M-class flare and followed by a few Type III bursts. The radio burst begins slightly before the X-ray flare. The GOES long-wavelength channel runs from 06:00 to 06:28~UT and the short-wavelength channel from 06:03 to 06:21~UT. The HEL1OS emission, which has a complex profile, runs from 06:03 to 06:18~UT, and its light curves peak before the GOES flux. The radio burst ends first, and the X-ray flare lasts about three times longer than the radio burst.

\section{Effects of SRBs}

As the Solar Radio bursts are the direct consequences of the Solar activities which influences the near-earth Space Weather. By detecting these bursts we can study the space weather and also predict the arrival of plasma from those events with the help of solar wind analysis. These prediction can help in avoiding the potential damages to satellites, astronauts, communication systems, and terrestrial power grids. Furthermore, the occurrence rate of SRBs is closely linked to the phase of the solar cycle, with an increased number of events generally observed during periods of enhanced solar activity around solar maximum.

%==================================
% IMAGE START
%==================================

%==================================
% IMAGE END
%==================================

\section{Future scopes and discussion}

Our observations on certain days indicate an enhancement in the overall solar radio intensity during periods of increased X-ray activity, even in the absence of clearly identifiable individual radio burst counterparts. This suggests that variations in the broadband radio emission may also provide useful information about solar activity. The existing data processing and analysis techniques could be further developed towards establishing a real-time solar radio monitoring system for space-weather applications. To facilitate the automated detection and identification of solar radio bursts, a machine learning based detection framework is currently under development.

Furthermore, establishing a permanent observational setup for routine solar monitoring would enable continuous tracking of solar radio activity. To gain further insight into the polarization characteristics of the detected events, polarization studies will be carried out for selected events using observations from DART Telescope \citep{DART} at 170 MHz. These observations may also provide an opportunity to investigate temporal variations in the solar magnetic field. In addition, the effects of solar events on solar wind properties will be investigated using the Solar Wind Ion Spectrometer (SWIS) and Solar wind Electron Proton Alpha Monitor (STEPS) data from the Aditya-L1 mission.\\

\bibliography{bibliography}

%==================================
% APPENDIX
%==================================

\clearpage
\onecolumn

\appendix
% jaa.cls puts \setcounter inside \thesection, which breaks \section here;
% same printed label, without the side effect.
\renewcommand\thesection{Appendix \Alph{section}.}

% Event list: 2023–2024. The section heading is typeset on this page, so it
% sits above the table instead of on a page of its own.
\includepdf[
    pages=-,
    fitpaper=true,
    pagecommand={\thispagestyle{plain}\section{List of Events Analysed}}
]{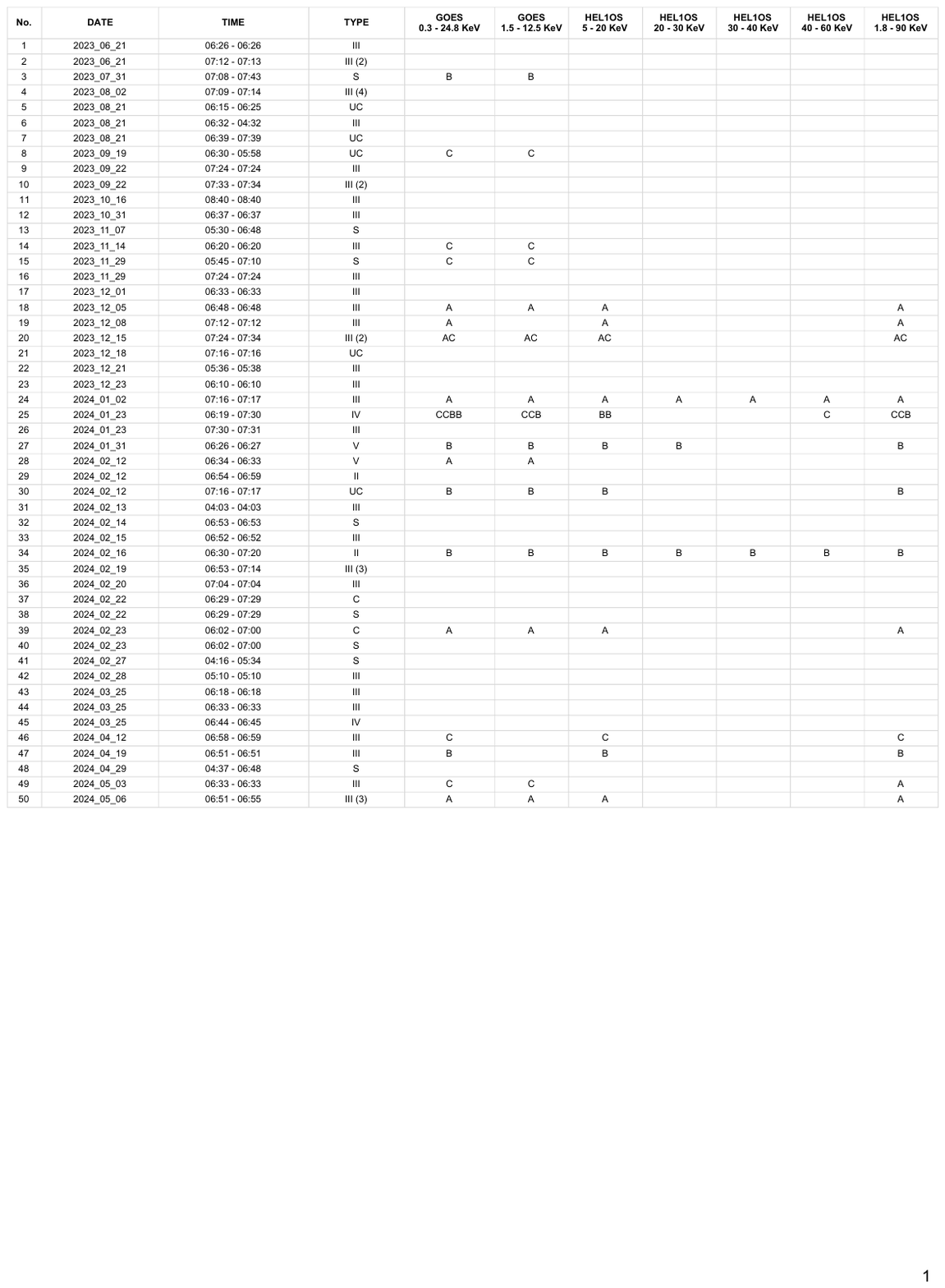}

% Event list: 2024–2025
\includepdf[
    pages=-,
    fitpaper=true,
    pagecommand={\thispagestyle{plain}}
]{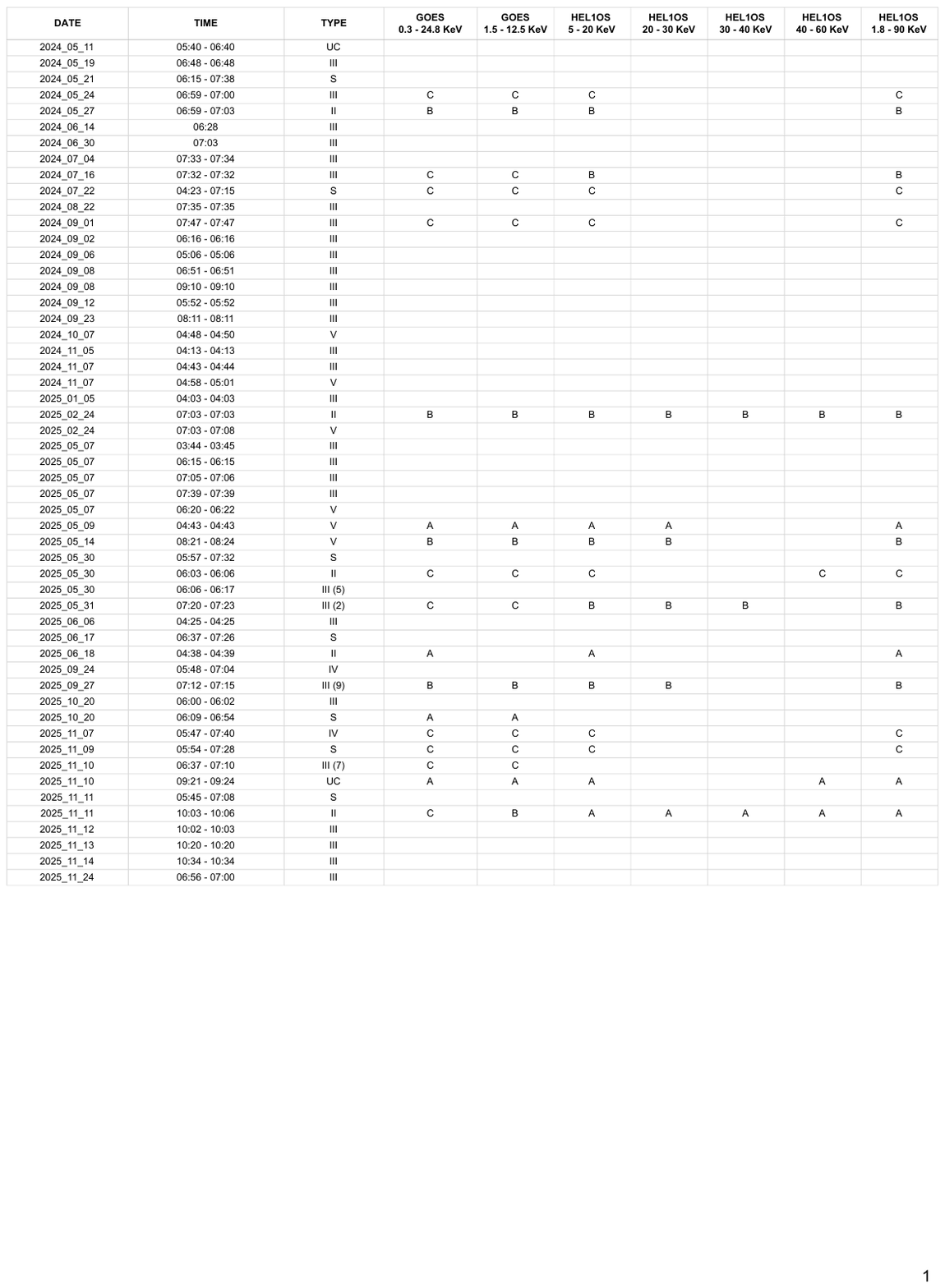}

\clearpage
\twocolumn

%==================================

%The table below includes the list of all the available

%\includepdf[pages=-]{event_list/twin_detections_comp_f1.pdf}

%\csvautolongtable{event_list/twin_detections_comp_f1.csv}

%\pgfplotstabletypeset[col sep=comma]{twin_detections.csv}

\end{document}